\documentclass[preprint,aps]{revtex4}

\usepackage{graphicx}
\usepackage{dcolumn}
\usepackage{bm}

\begin{document}


\title{Heat capacity of a  two-component superfluid Fermi gas}

\author{Alexander V. Avdeenkov}
 \altaffiliation[JILA ]{Physics Department, University of Colorado.}
 \email{avdeyenk@murphy.colorado.edu}
\affiliation{%
JILA, University of Colorado
}%


\date{\today}

\begin{abstract}
We investigate mean-field effects in two- component
trapped Fermi gases in  the superfluid phase, in the vicinity of
 $s$-wave Feshbach resonances.  Within the resonance superfluidity approach~\cite{holland1} we calculate
the ground state energy and the heat capacity as function of temperature.
Heat capacity is analyzed for different trap aspect ratios. We find that trap anisotropy is an important factor in determining both the
value of  heat capacity near the transition temperature and  the transition temperature itself.
\end{abstract}

\maketitle


At present several groups~\cite{regal,thomas,bourdel} have produced Fermi gases of atoms at temperatures where the superfluid
 phase is expected.
The possibility of tuning interatomic  interactions near Feshbach resonances may provide  a chance to
watch  macroscopic phenomena for both normal and superfluid phases
and the phase transition from one to another. 
The presently reachable temperatures limit the existence of the superfluid  phase to a strongly- interacting gas.
 For now  it is still an open question how to observe the superfluid transition,
although numerous proposals exist~\cite{nygaard,bulgac,bruun,torma,menotti,cozzini}(and references therein).

 Here we examine the thermodynamics
of a two-component Fermi gas in a strongly- interacting regime where we can expect  superfluidity.
Although any attractive interaction can theoretically  support Cooper pairing at zero temperature, 
  it seems that
only a strongly- attractive interaction may produce observable effects in current experiments. For a strongly- interacting gas
the 'small parameter' $k_{F}a>1$ and so the ordinary mean-field approach breaks down. 
In this case the resonance superfluidity \cite{holland1,holland, griffin} approach suits the problem best.
On the experimental side the strongly- interacting regime is reached by exploiting  magnetic-field Feshbach 
resonances~\cite{regal,thomas,bourdel}.
 For our numerical simulation we took the parameters of the  two-
component Fermi gas of $^{40}K$  near Feshbach resonance.  These
components are in $|9/2-9/2>$ and $|9/2-5/2>$ states. In
\cite{regal} it was shown that there is an s- wave Feshbach resonance  at a
magnetic field $224.21 \pm 0.05$G and near this resonance the scattering length was
measured up to about $\pm 2000a_0$.

 Kokkelmans et al~\cite{holland} employed the resonance
 superfluidity approach in order to get the critical temperature and find possible signs
 of the phase transition. Ohashi and Griffin \cite{griffin} independently used the same approach.
For the superfluid phase we use  the mean-field approach of BCS-type theory within the local density approximation
and use the Hamiltonian of the resonance superfluidity  model \cite{holland} in order to take Feshbach resonance
physics into account.
 In this case the energy density functional ${\cal E}$ depends on normal  $\rho({\bf x,y})$,
anomalous $\kappa({\bf x,y})$ densities and a molecular field $\phi(\frac{{\bf x}+{\bf y}}{2})$ 
where ${\bf x}$ and ${\bf y}$ are coordinates of two atoms. The set of equations
as well as their solution for energy- independent effective interaction(see~\cite{ring} for example)
 look like in ordinary Bogolyubov-de Gennes formulation but with the extra term of a molecular field:
\begin{eqnarray}
\label{hfb}
\nonumber
\rho({\bf k,R})=n({\bf k,R})u^{2}({\bf k,R})+
(1-n({\bf k,R}))v^{2}({\bf k,R}), 
\\
\nonumber
\kappa({\bf k,R})=u({\bf k,R})v({\bf k,R})(1-2n({\bf k,R}))
\\
\nonumber
E({\bf k,R})=\sqrt{h({\bf k,R})^2+\Delta({\bf R})^2}, 
\\
\nonumber
{u^{2}({\bf k,R}) \choose v^{2}({\bf k,R})}=\frac{1}{2}(1 \pm \frac{h({\bf k,R})}{E({\bf k,R})})
\\
h({\bf k,R})=\frac{\hbar^2 k^2}{2m} + V_{mf}({\bf R})+V_{trap}({\bf R}) - \lambda 
\\
\nonumber
 V_{mf}({\bf R})=V_{bg}\rho({\bf R});\hspace{1cm}
\rho({\bf R})=\int \frac{d^{3}k}{(2\pi)^3}\rho({ \bf k,R})
\\
\nonumber
\Delta({\bf R})=-V_{bg}\int \frac{d^{3}k}{(2\pi)^3}\kappa({\bf k,R})-g\phi({\bf R})
\\
\nonumber
\phi({\bf R})=\frac{g}{2\lambda-V_{trap}^{mol}({\bf R})-\nu} \int \frac{d^{3}k}{(2\pi)^3}\kappa({\bf k,R})
\end{eqnarray}
where ${ \bf k}$ is a wavevektor, ${\bf R }= \frac{{\bf x}+{\bf y}}{2}$ is the coordinate of the  center of the mass for two
interacting atoms, $n({\bf k,R})=(exp(E({\bf k,R})/k_{B}T)+1)^{-1}$ is the Fermi-Dirac distribution, 
 $V_{mf}({\bf R})$ is a mean field potential and $\Delta({\bf R})$ is the energy gap.
$V_{bg}=4 \pi \hbar^2 a_{bg}/m$, 
$a_{bg}$ is the background scattering length, $g=\sqrt{V_{bg} \Delta B \Delta \mu}$ is the coupling strength
 and, $\nu=(B-B_{0})\Delta \mu $ is the magnetic field detuning, $\Delta B$ is the field width of the resonance, $\Delta \mu $ is the 
magnetic moment difference between two hyper- fine levels of the  two-component Fermi gas~\cite{holland,timmermans}.
More complete theoretical analysis  using this approach can be found in~\cite{griffin}.
Functions $f(k,R)$ are the Wigner transforms of corresponding functions $f(x,y)$.
So after solving (\ref{hfb}) we have both a normal $\rho({\bf k,R})$ and an anomalous $\kappa({\bf k,R})$  distribution function as well as
the molecular field $\phi ({\bf R})$
and we  find the ground state energy as:
\begin{eqnarray}
\label{gse}
\nonumber
E=\int \frac{d^{3}k}{(2\pi)^3}d^3 R \{ k^{2} \rho({\bf k,R}) +V_{bg}\rho({\bf k,R}) \rho({\bf R})
\\
+\rho({\bf k,R})2V_{trap}({\bf R})
- \Delta({\bf R}) \kappa^{*}({\bf k,R}) \}
\\
\nonumber
+\int d^3 R V_{trap}^{mol}({\bf R})\phi^{2}({\bf R})
\end{eqnarray}

It is a  well- known phenomenon that near the phase transition the energy-temperature curve should have a distinct change.
Moreover we can investigate  the dependence of specific heat capacity  and  ground state energy  on  interaction 
strength and 
trap aspect ratio. As we will see the geometry of the trap strongly influences  the thermodynamics of the superfluid gas.

We consider an anisotropic trap with  trap aspect ratio $\lambda=\omega_{z}/ \omega_{\perp}$ of the transverse and axial
frequencies. Because the ratio $\lambda$ is rather small in the current experiments~($\approx 0.01$)~ \cite{regal,thomas} 
calculations for isotropic or
almost isotropic traps possess only a methodological interest.
However, it is in nearly isotropic traps that the most dramatic observables are expected; see below.
 As the interaction near a Feshbach resonance strongly depends on
detuning we consider  all the characteristics of this system  as a function of detuning $\nu$ and temperature $T$.
For present calculations we chose $\omega_{\perp}=400Hz$ and $5 \times 10^5$ particles. 
As the detuning is defined  by the magnitude of a magnetic field  we will report  detuning in units of Gauss, which is more
convenient for a  possible comparison with  experiment.

\begin{figure}
\newpage
\centerline{\includegraphics[width=0.9\linewidth,height=1.08\linewidth,angle=-90]{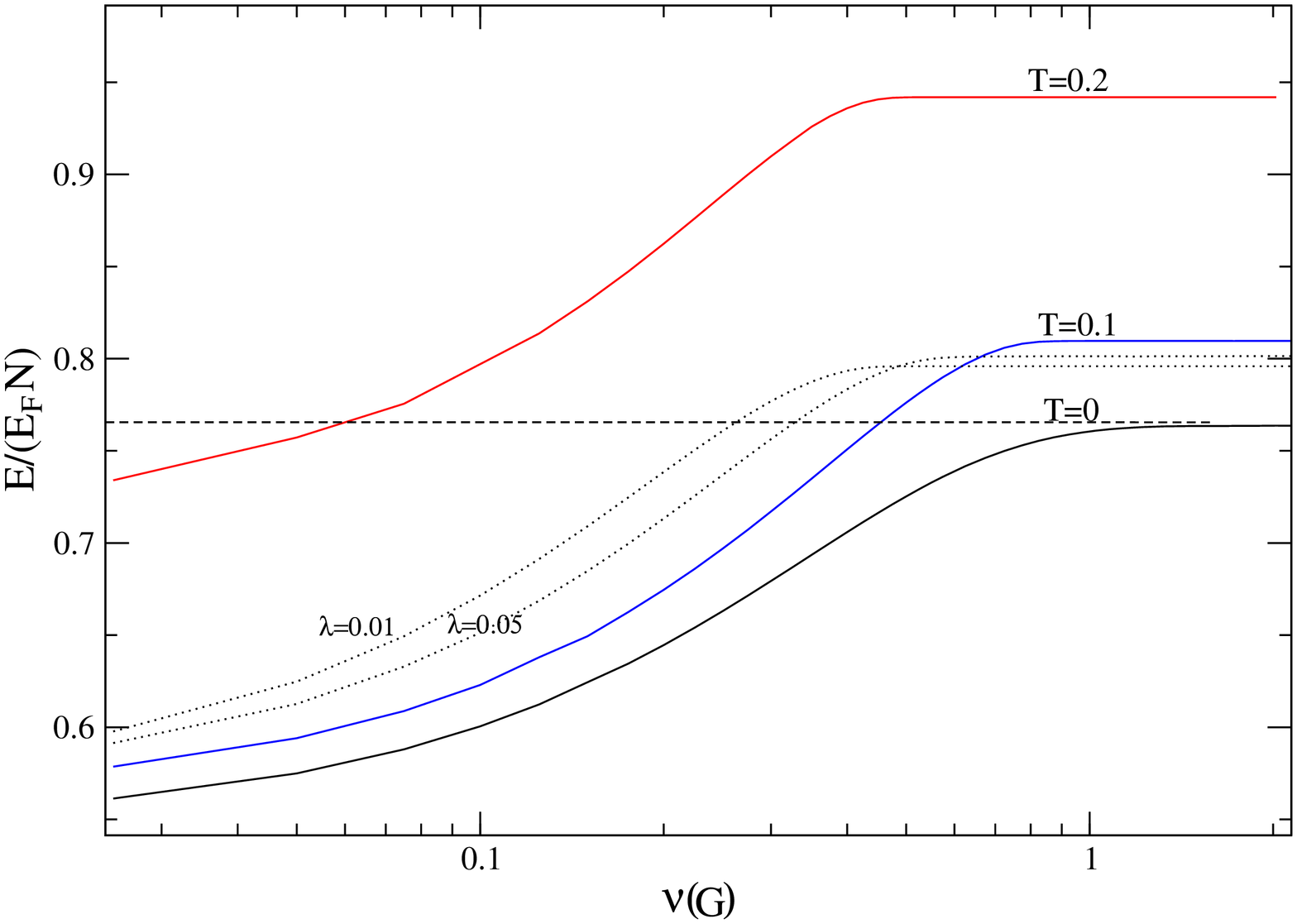}}
\caption{Ground state energy per particle as a function of the detuning for different temperatures.
The temperature is in $T_{F}$ units. The trap aspect ratio $\lambda =0.5$ for the solid curves.
The dotted curves display different aspect ratios for $T=0.1T_{F}$
}
\label{gsed}
\end{figure}

\begin{figure}
\centerline{\includegraphics[width=0.9\linewidth,height=1.08\linewidth,angle=-90]{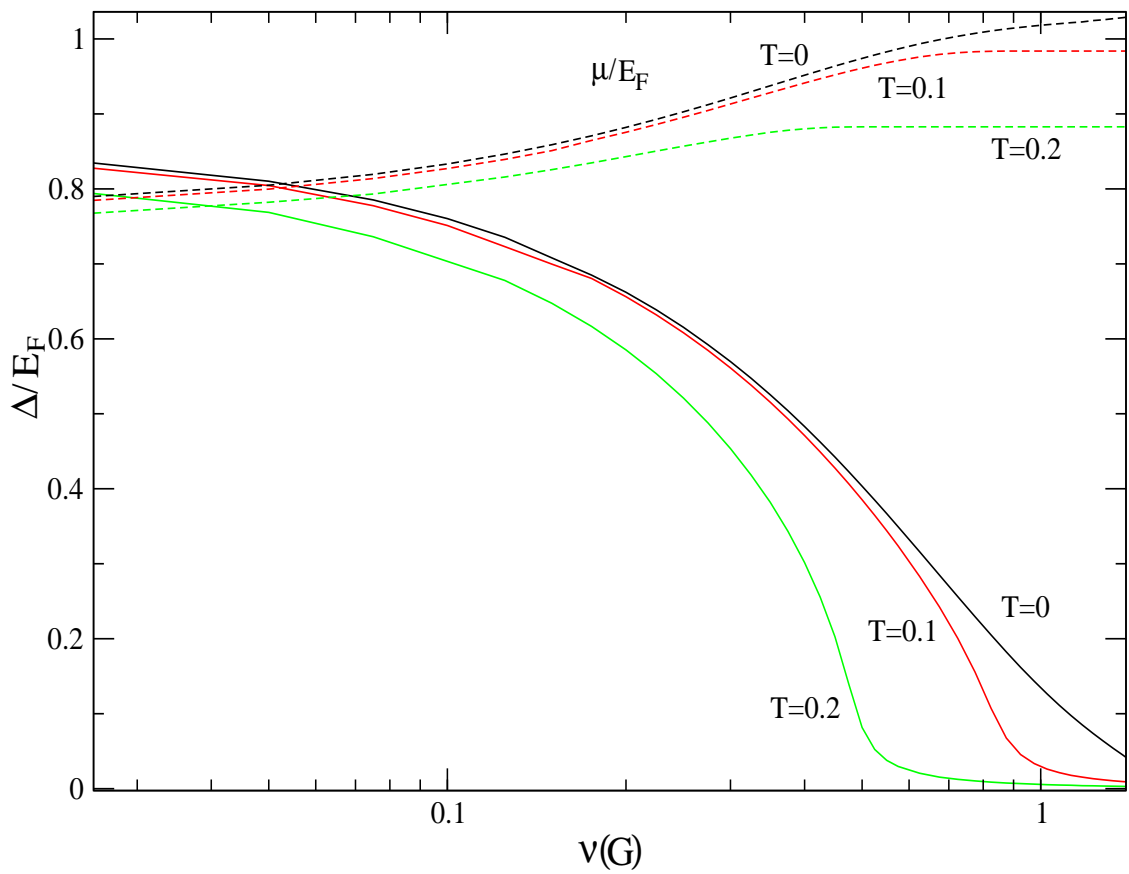}}
\caption{Energy gap in the center of the trap and chemical potential as a function of the detuning for different 
temperatures. $\lambda =0.5$.}
\label{gapd}
\end{figure}

Within this approach we have calculated the ground state energy and  heat capacity for a variety
of interactions and temperatures.
The typical dependence of ground state energy and gap on detuning is shown in~Fig.( \ \ref{gsed}, \ref{gapd}).
In the range of  detuning shown the interaction is always attractive so a superfluid phase can exist. But at a detuning
larger than $\approx 1G$ the pairing energy seems rather small compare to the kinetic energy as the
energy- detuning dependence is very weak. In the vicinity of a resonance the mean-field $V_{mf}$ contribution to
the ground state energy  is negligible as the background part of the scattering length is small compared to
the resonance part.
With decreasing  detuning the pairing energy  becomes a significant part of the total energy and we can see considerable 
lowering of the total energy beginning  from some detuning that depends on temperature too. 
Unfortunately it is impossible to
estimate the kinetic and the pairing energies separately because the integral of the second and forth terms in~(\ref{gse}) 
over momentum are individually divergent~\cite{bertch}.

 Fig.\ \ref{gsed} thus shows that  detunings smaller than $\approx 1G$
seem required to observe superfluidity.
In  Fig.\ \ref{gsed}  the solid curves are calculations for
$\lambda=0.5$. The dotted curves are for $\lambda =0.05; 0.01$ and for $T=0.1T_{F}$.
So for more anisotropic traps the pairing energy is smaller.
 It should be possible to measure the energy, as in~\cite{jin}. In~\cite{jin}
the released  energy was measured for Bose gas above and below the transition temperature. The uncertainty is
 $\approx 10-30 \%$ for temperatures below $T_{c}$. 
The desired detuning should therefore be small enough to generate at least a $10 \%$ change in energy compared to  the normal phase energy.
Thus
the detuning should be  smaller than $0.5G$ for $T/T_{F}=0.1$, $\lambda = 0.5$.
For more realistic trap aspect ratios (0.01) it should be smaller than $0.3G$ 
  At these values of detuning the pairing energy is not a small part of the total ground state energy which means that
the energy gap is  rather large too (Fig.\ \ref{gapd}).

From experimental point of view  it seems that the closer to the resonance we are the more chances to observe
superfluidity we have.
But it should be mentioned that  near a resonance there is so called  the BCS-BEC crossover  regime~\cite{sademelo}.
In theory of the superconductivity at this regime there is a class of 'exotic', high-$T_{c}$ superconductors. 
 In the case
of the crossover regime ($\mu / \Delta \leq 1$) we need a more sophisticated theoretical~\cite{milstein} approach 
than the ordinary mean- field theory.  Such a treatment is outside of scope of the present article, but it may be important
in order to get
the right value of the critical temperature~\cite{haussmann} for this regime.
If $\mu / \Delta$ is sufficiently greater than $1$ we can hope we have the ordinary BCS regime. 
For smaller trap ratios the condition  ($\mu = \Delta $) happens at a slightly smaller detunings. 

\begin{figure}
\centerline{\includegraphics[width=0.9\linewidth,height=1.08\linewidth,angle=-90]{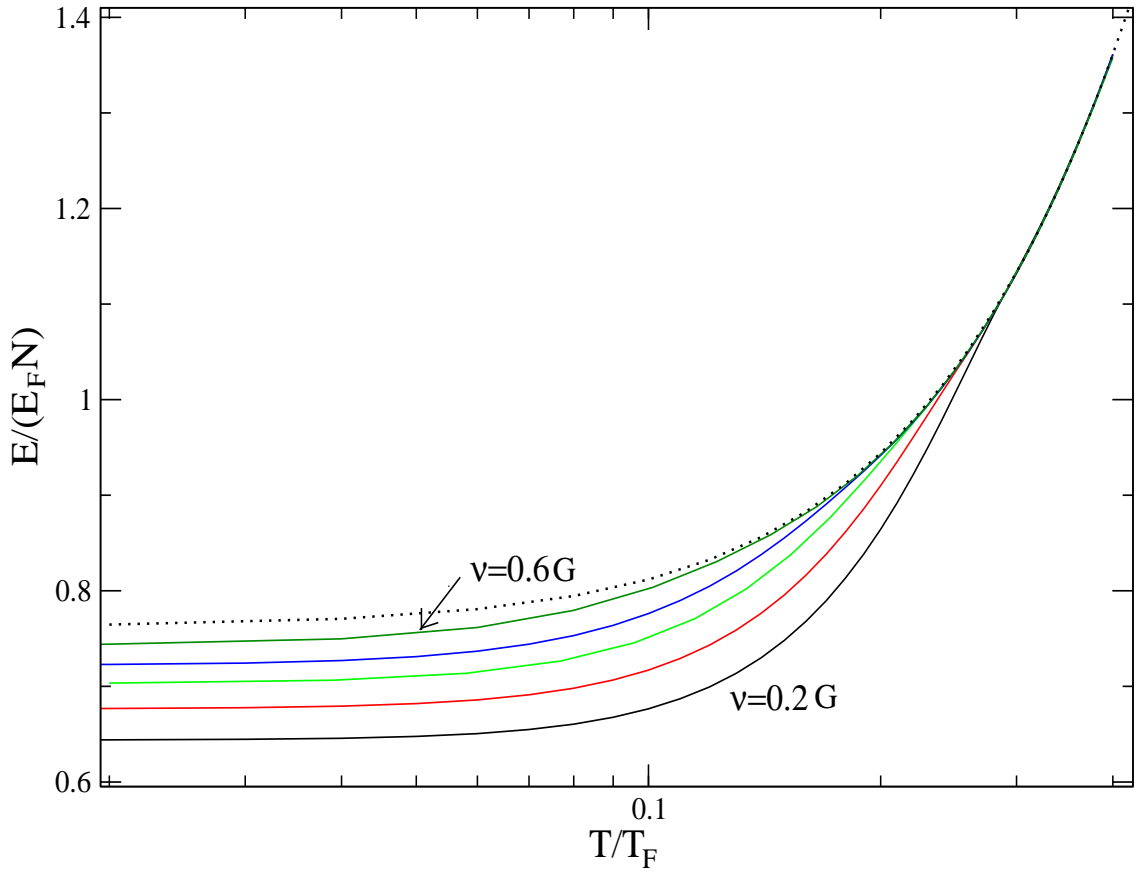}}
\caption{Ground state energy per particle as a function of a temperature for
the magnetic field detuning $\nu=0.2-0.6G$ with $step=0.1G$
The dotted line is the energy in the case when the interaction is described just by the non- resonant scattering length.
}
\label{gset}
\end{figure}

In this article we do not
analyze  how much larger $\mu / \Delta $ should be for the theory to hold,  and leave such an analysis for the near future.
So for further investigations we have chosen detunings $\nu=0.3-0.5G$  where the energy gap is still significant.
The  energy-temperature dependence  (Fig.\ \ref{gset}) is a more appropriate characteristic for the comparison and 
describing of  experimental data.
Again from an experimental uncertainty point of view it seems that temperatures 
around $0.1T_{F}$ or smaller are  more appropriate to observe  superfluidity.
For chosen parameters of interest we have the condition $\Delta \gg \hbar \omega$ and according
to the pairing classification of~\cite{bruun2} this is an inter-shell pairing regime for which
 the local density approximation is the appropriate tool for describing the pairing.

\begin{figure}
\centerline{\includegraphics[width=0.9\linewidth,height=1.08\linewidth,angle=-90]{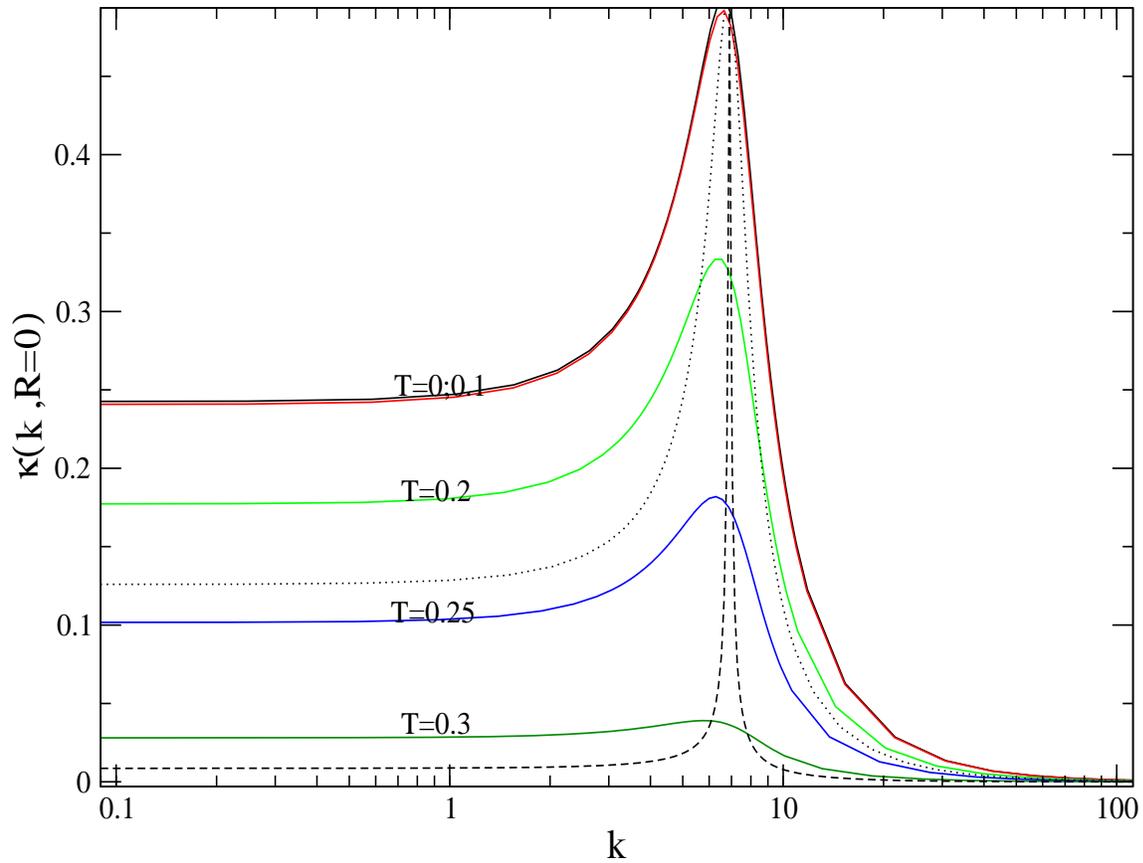}}
\caption{Distribution function  $\kappa({\bf k,0})$ in the center of the trap versus momentum for the pairing field for different temperatures in the case  $\nu =0.2G$
and $\lambda =0.01$. Also shown are the cases  for $\nu =0.4$(dotted line) and $\nu= 1G$(dashed line) and for
$T=0.1T_{F}$.
}
\label{kappa}
\end{figure}

The important characteristic of the pairing field is its distribution in momentum $\kappa({\bf k,R})$~(Fig.\ \ref{kappa}). 
In the same figure we demonstrate this distribution for $\nu=0.4;1 G$~(dashed and dotted curves) for $T=0.1T_{F}$. 
 Far from resonance~($\nu =1G$, dashed line) this function  has just  a narrow peak
near Fermi momentum $k_{F}$.
 With decreasing detuning more and more atoms  below $k_{F}$ are involved in the pairing  which emerges not only as 
a Fermi surface effect.
 As it was shown in \cite{pistolesi} the product  $k_{F} \xi $ ($ \xi $-
the coherence length) is the appropriate variable  for 'exotic' superconductors. Moreover the value $ \xi $ can be
considered as the size of the Cooper pair.
Within our approach the coherence length is  dependent on location 
$\xi^{2}({ \bf R})=\int d{\bf r} \kappa({\bf r,R}) r^{2} / \int d{\bf r} \kappa({\bf r,R}) 
\approx (k_{F}({\bf R})/ m \pi \Delta({\bf R}))^2 $. We have calculated this for the center of the trap and found that 
$\xi$ is about 0.33-1.65 in trap units for $\nu = 0.3-0.5G$ and $\lambda =0.01$
But the interparticle distance is changing much slower and is around $0.35$ for the given detunings.
 For our reasonable detunings and $\lambda=0.01$ the parameter  $k_{F} \xi $ is
2.26-8 but the `boundary'  which distinguishes the high-$T_{c}$ and conventional superconductors~\cite{pistolesi}
 is $k_{F} \xi \approx 2 \pi $. It was suggested~\cite{uemura} that the `exotic' superconductors are intermediate between
BCS- type supeconductors and BEC. Holland el al~\cite{holland1} predicted the same phenomenon for the two-component 
degenerate Fermi gas. So  according to this classification we have conventional superfluidity for $\nu \approx 0.5G$
and larger. Moreover at this field $k_{F}a < 1$  and the ordinary mean-field theory can be used.

The coherence length can give us an insight into whether it is possible to detect a signal   
of the Cooper pair breaking, similar to what was done
for bound molecular states at negative magnetic- field detunings~\cite{regal2}. If it is possible, this can be considered as a
sign of superfluidity. From this point of view it seems that it is not desirable to be very close to resonance
because in this case  the many-body wave function of the Cooper pair at distances of the
coherent length order 
between atoms will be very close to the two- body wave function for quasi- bound states near a resonance.
At large distances between atoms the wave function of the Cooper pair  will be very different 
from the relative wave function of two scattering atoms but it is nor clear if
the  RF spectroscopy can  work at such distances. Our approach enables us to find the Cooper pair wave function and
in the future  we will analyze this aspect in more detail.

The phase transition to  superfluidity stipulates a considerable increase in heat capacity near the critical 
temperature.  This 
characteristic may therefore  serve as an observable sign of superfluidity.
The main point of the present article is an investigation of a heat capacity for Fermi gas for the conditions  
described 
above. In~\cite{bruun3} the authors calculated and analyzed  this characteristic for off- resonance interacting $^{6}Li$ 
within 
ordinary HFB theory and  the effect is too small  to be detected in current experiments. But the resonance superfluidity 
approach gives us a chance  to do this analysis for strongly- interacting systems. 
The phenomenon of superfluidity strongly depends on the trap geometry.
 For an s-wave interacting gas  the ${\bf R}-$ dependence 
on the energy  gap~(\ref{hfb}) reflects only the trap geometry . In our calculations we keep  $\omega_{\perp}=400Hz$ and 
vary   $\omega_{z}$.  It is clear that with decreasing $\lambda$ the energy gap will be smaller in the center of the 
trap  than for an isotropic trap. So the spatial manifestation of the superfluid 
phase~\cite{holland} depends on geometry too.

\begin{figure}
\centerline{\includegraphics[width=0.9\linewidth,height=1.08\linewidth,angle=-90]{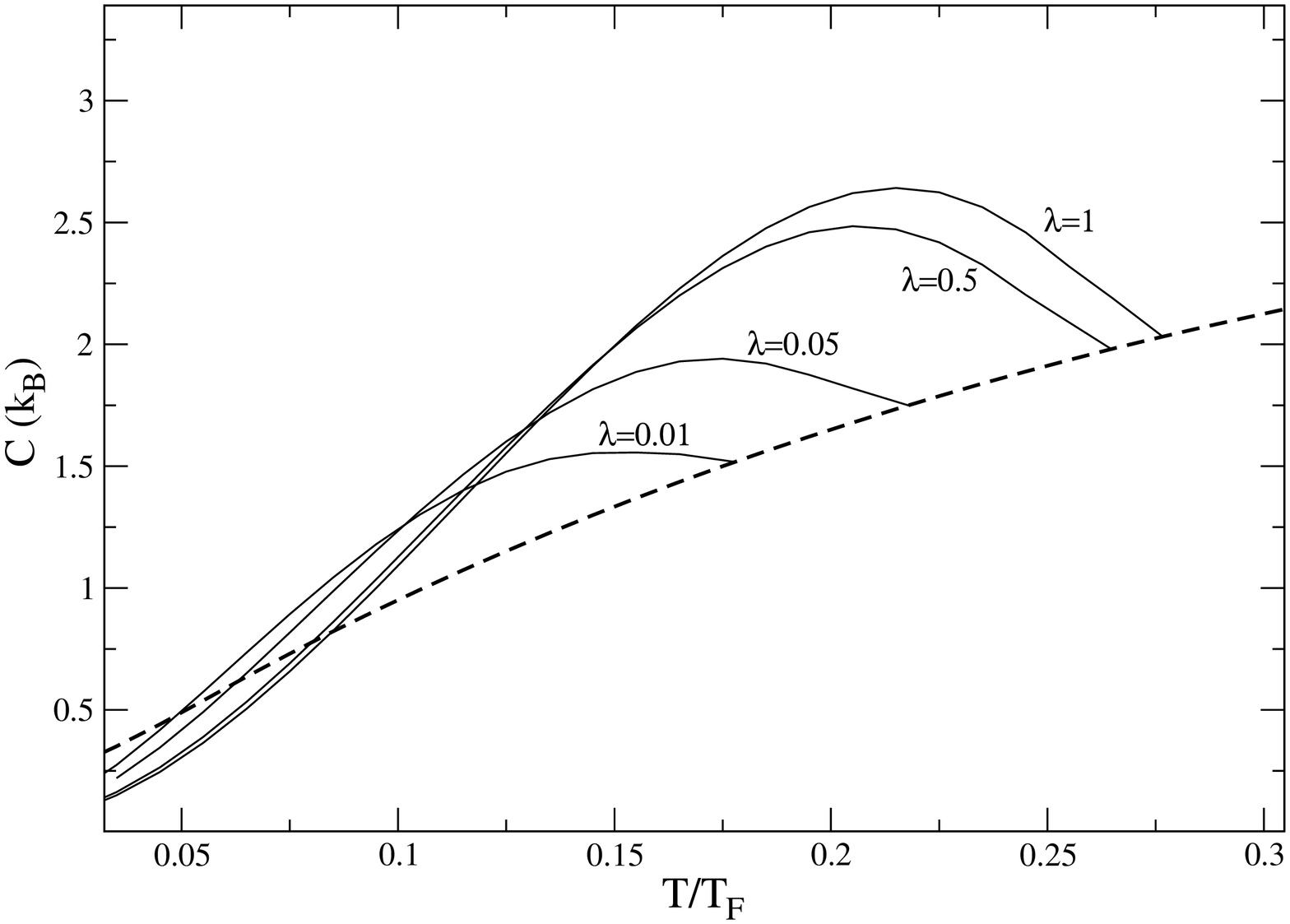}}
\caption{Heat capacity as a function of a temperature for  detuning $\nu=0.3G$ and for different
trap aspect ratios $\lambda$. The dashed line is for the normal phase gas.
}
\label{heat1}
\end{figure}

\begin{figure}
\centerline{\includegraphics[width=0.9\linewidth,height=1.08\linewidth,angle=-90]{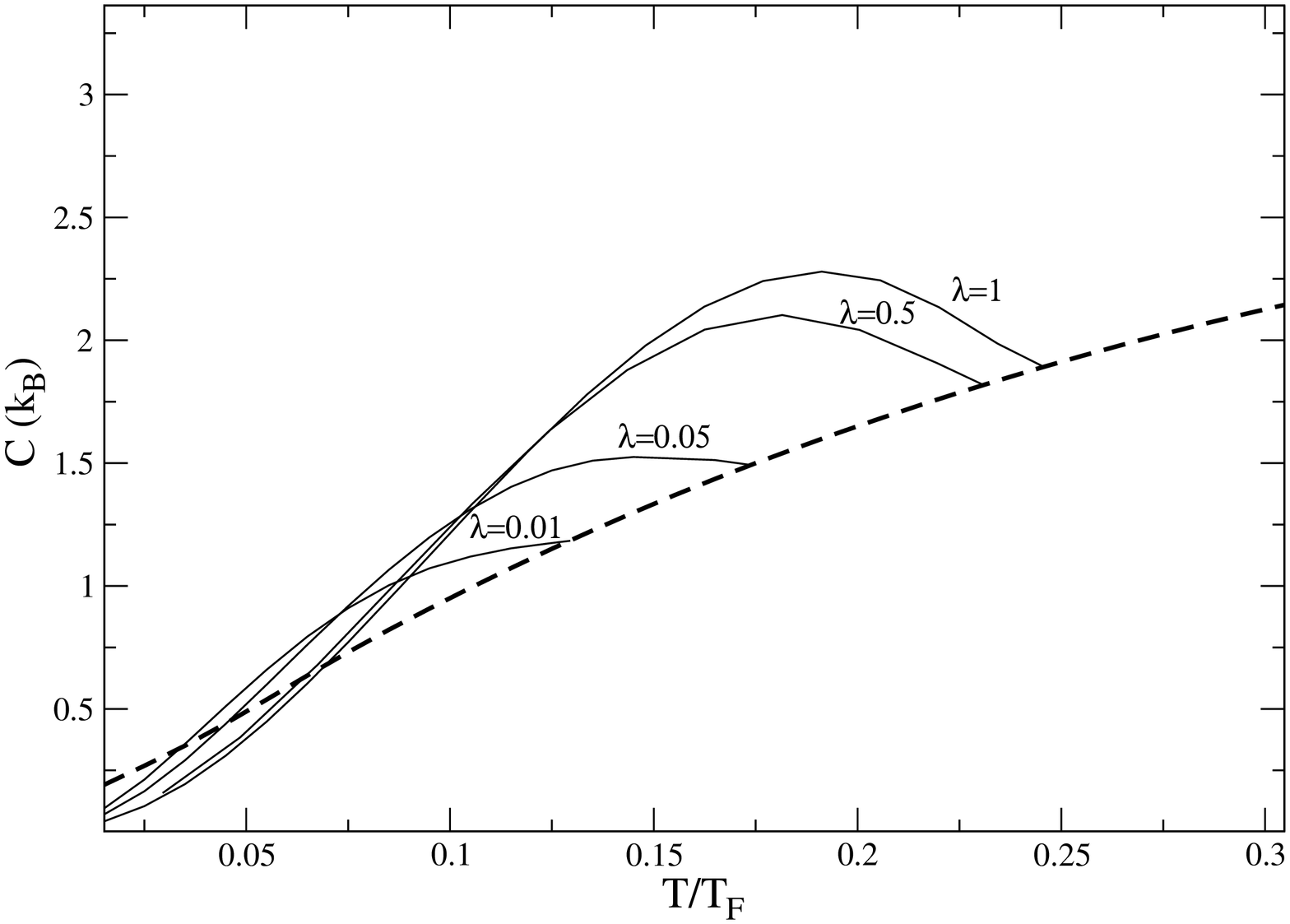}}
\caption{Heat capacity as a function of a temperature for  detuning $\nu=0.4G$ and for different
trap aspect ratios $\lambda$.
The dashed line is for the normal phase gas.
}
\label{heat2}
\end{figure}

 We found  (Fig.\ \ref{heat1}, \ref{heat2})  that the trap anisotropy considerably 
influences the heat capacity of a gas in the superfluid phase  and can wash out the  effect altogether
 if  $\lambda$ is very 
small.
It is known that at the critical temperature the heat capacity has a discontinuity.
In the trap-confined gas the $T_{c}$ as well as the gap is $\bf R$- dependent and the higher the temperature is, the
smaller is the fraction  of the gas  involved in superfluidity. The critical temperature is the point at which the energy gap
disappears in the center of the trap. Above this temperature the gas is in the normal phase everywhere.
 We can see  that for temperatures lower
then $T_{c}$ heat capacity of the superfluid phase can be  up to $\approx 1.5$ times larger than for  normal phase gas, and 
this value depends on
the trap aspect ratio and detuning.
Also the more anisotropic a trap is the smaller the gap is and then smaller a 'bump' in heat capacity for superfluid
phase. 

The heat 
capacity as well as the energy of ground state can be measured using a ballistic expansion  as  was done for a
Bose gas~\cite{jin}. But, as it was pointed out in~\cite{regal,thomas} an expanding strongly- interacting gas 
is likely in the hydrodynamic regime.
 We have estimated that for temperatures near the critical temperature and at small detunings~(0.2-0.4G) the 
collision rate~\cite{regal,thomas} can be larger than the trap frequency, which also depends on the trap aspect ratio.
It means that  the hydrodynamic regime is very probable during the expansion.
We can choose the trap aspect ratio in order to get the collision rate a bit smaller than the transverse frequency. 
For $\nu=0.3G$ at the corresponding critical temperature the ratio $\lambda$ should be around  0.01(Fig.\ \ref{heat1}).
A similar estimation  for $\nu=0.4G$ shows that $\lambda$ can be around 0.05(Fig.\ \ref{heat2}).
 It is clear that for reasonable parameters near the resonance  the cloud will not be  expanded ballistically as a
 whole. However, a measurement of  the total energy  after turning-off the trap is still valid, since
the released energy should be conserved. 
The trap energy is
almost two times smaller than total energy so the energy- temperature dependence  of the released energy  will be very 
similar to that shown in Fig.\ \ref{gset} but approximately two times smaller.

This work was supported by the NSF. The author appreciates J.L.Bohn and M.Holland for stimulating and useful discussions 
and C.Ticknor for providing with  parameters of the resonance for $^{40}K$.

\end{document}